\documentclass{article}

\usepackage{amsmath,amsfonts,amssymb}
\usepackage{graphicx}
\usepackage{caption}
\usepackage{subcaption}
\usepackage[top=2cm, left=2cm, bottom=2cm, right=2cm]{geometry}
\usepackage{epstopdf}
\usepackage{hyperref}
\usepackage{color}
\usepackage{cite}
\usepackage[utf8]{inputenc}

\begin{document}

\title{Generalized revival and splitting of an arbitrary optical field in GRIN media}

\author{H.M. Moya-Cessa$^{1}$, F. Soto-Eguibar$^{1,*}$, V. Arrizon$^1$ and A. Z\'u\~niga-Segundo$^{2}$,  \\
$^1${\small Instituto Nacional de Astrof\'{\i}sica, \'Optica y Electr\'onica, INAOE} \\
\small {Calle Luis Enrique Erro 1, Santa Mar\'{\i}a Tonantzintla, Puebla, 72840 Mexico}\\
$^2${\small Departamento de F\'{\i}sica, Escuela Superior de F\'{\i}sica y Matem\'aticas, IPN},\\
 {\small Edificio 9, Unidad Profesional Adolfo L\'opez Mateos, DF, 07738 Mexico} \\
$^*$\small {Corresponding author: feguibar@inaoep.mx}}

\date{\today}

\maketitle

\begin{abstract}
Assuming a non-paraxial propagation operator, we study the propagation of an electromagnetic field with an arbitrary initial condition in a quadratic GRIN medium. We show that at certain specific periodic distances, the propagated field is given  by the fractional Fourier transform of a superposition of the initial field and of a reflected version of it.  We also prove that for particular wavelengths, there is a revival and a splitting of the initial field. We apply this results, first to an initial field given by a Bessel function and show that it splits into two generalized Bessel functions, and second, to an Airy function. In both cases our results are compared with the exact numerical ones.
\end{abstract}


\section{Introduction}
Graded index (GRIN) media are mainly used in image formation applications \cite{gomez}. On the other hand, it has been established that a GRIN medium can form self-images of periodic fields  \cite{R1,R2}  and can support invariant propagation modes, either in the paraxial \cite{ozaktas} and the non-paraxial domains \cite{ojeda}. In a different context, light propagation in a quadratic GRIN medium can be employed as a form of optical emulation of quantum phenomena. An example is the mimicking of quantum mechanical invariants by the propagation of light through the interface of two coupled GRIN devices \cite{Arr}. Because the Schr\"odinger equation and the paraxial wave equation in classical optics are formally equivalent, cross applications between quantum mechanics and classical optics are common. One can extend the application in order to consider not only the paraxial regime but also the non-paraxial one, i.e. the complete Helmholtz equation. For instance, supersymmetric methods, common to quantum mechanics, have been proposed in classical optics \cite{Chuma,David}. 

An interesting conceptual and mathematical result is that the paraxial propagated field in a quadratic GRIN medium can be expressed as the Fractional Fourier transform (FrFT) of the incoming beam \cite{ozaktas}. Indeed, the FrFT operator has been considered for discussing additional similitudes between classical optics and quantum formalism.  For instance, Agarwal and Simon \cite{Agarwal} have shown that Fresnel diffraction leads to the FrFT by noting that, when constructing the (quantum) harmonic oscillator evolution operator, it contains a term proportional to paraxial free propagation. In another report, by Fan and Chen \cite{Fan},  it is shown that the quantum-mechanical position-momentum mutual transformation operator is the core element for constructing the integration kernel of FrFT.

In the context of GRIN media, with quadratic refractive index dependence in the radial coordinate, knowledge from harmonic oscillator-type Hamiltonians can be used for the solution beyond the paraxial regime. As a significant result in this context, we established previously the long period revival and splitting of a Gaussian beam, transmitted by a quadratic GRIN medium \cite{Eguibar,Arrizon}. Such effects are theoretically predicted expressing the Taylor series for the propagation operator up to the second order, i. e. including an additional term beyond the paraxial approximation. 

In the present paper, we describe a significant generalization of such long period effects, assuming again a second order form of the propagator. We establish propagation distances that exhibit the revival and the splitting of an arbitrary input field. We also establish propagation distances for which any input field $E(x)$  splits into two fields given by the FrFT of $E(x)+E(-x)$. We apply this result to an initial field given by a Bessel function, and as a solution, we obtain the superposition of two so-called Generalized Bessel functions \cite{Dattoli,Torre,Dattoli2,LeijaI,LeijaII,Torre2}.

\section{Helmholtz equation for GRIN media}
The Helmholtz equation in two dimensions for a GRIN medium is
\begin{equation}
\left[    \frac{\partial^2}{\partial X^2}+\frac{\partial^2}{\partial Z^2}+\kappa^2 n^2(X) \right]  E(X,Z)=0,
\end{equation}
where $\kappa$ is the wave number and $n(X)$ is the variable refraction index.
For a quadratic medium, the refraction index can be written as
\begin{equation}
	n^2(X)=n_0^2\left( 1-g^2 X^2\right),	
\end{equation}
where $g$ is the gradient index in the $X$ direction. So, for a quadratic dependence in the index of refraction, the Helmholtz equation is expressed as
\begin{equation}
\left(     \frac{\partial^2}{\partial z^2} +\frac{\partial^2}{\partial x^2}+k^2- x^2 \right) E(x,z)=0,
\end{equation}
where we have introduced the dimensionless variables $x=\sqrt{\eta}X$ and $z=\sqrt{\eta}Z$, and where we have defined $\eta= n_0 g \kappa$ and $k= n_0 \kappa/\sqrt{\eta}$.\\
Introducing the operator $\hat{p}=-i\frac{d}{dx}$, we can cast the last expression as \cite{wolf}
\begin{equation}
\frac{\partial^2 E(x,z)}{\partial z^2}= -\left(k^2-\hat{p}^2- x^2 \right)E(x,z), 
\end{equation}
whose formal solution is
\begin{equation}\label{040}
E(x,z)=\exp\left[-i z \sqrt{k^2-\hat{p}^2-x^2}\right] E(x,0)
=\exp\left[-i z k \sqrt{1-\frac{\hat{p}^2+ x^2}{k^2}}\right] E(x,0),
\end{equation}
where $E(x,0)$ is the boundary condition for $z=0$.\\
We define the lowering ladder operator, $\hat{a}=   (1/2)^{1/2}    \left( x+i\hat{p} \right)$, its adjoint, the raising ladder operator, $\hat{a}^\dagger=(1/2)^{1/2}  \left( x-i\hat{p} \right) $ and the number operator $\hat{n}=\hat{a}^\dagger\hat{a}$, and we write Eq. \eqref{040} as
\begin{equation}\label{050}
E(x,z)=\exp\left[-i z k \sqrt{1-\frac{2\hat{n}+1}{k^2}}\right] E(x,0),
\end{equation}

\subsection{The paraxial approximation.}
As a background for our main result in next section, we present here an alternate derivation of the paraxial propagation  in GRIN media, in terms of the FrFT \cite{Fan,pellat}. The paraxial approximation is obtained when the square root in the exponential of Eq. (\ref{050}) is expanded to first order, that is
\begin{equation}\label{060}
E(x,z)=\exp\left[ -i z k \left(1-\frac{1}{2k^2} \right)  \right]
\exp\left( i \frac{z}{k} \hat{n}\right)   E(x,0).
\end{equation}
On the other hand, it has been established that the fractional Fourier transform of a well behaved function $f(x)$ can be obtained in terms of the number operator $\hat{n}$, as
$ \mathfrak{F}_\alpha \left\lbrace f(x) \right\rbrace =\exp\left(i \alpha \hat{n} \right) \left\lbrace  f(x)\right\rbrace$, where alpha is the transform order. Considering this result, \eqref{060} can be rewritten as
\begin{equation}
E(x,z)=\exp\left[- iz\left( k-\frac{1}{2k}\right) \right] 
\mathfrak{F}_{\frac{z}{k}} \left\lbrace E(x,0) \right\rbrace.
\end{equation}
Thus, the paraxial propagation to a distance $z$ is proportional to the fractional Fourier transform of order $\frac{z}{\pi}$ of the initial condition $E(x,0)$ \cite{Fan,pellat}.\\

\subsection{Beyond the paraxial approximation}
We now allow ourselves to go one step further than the paraxial approximation. In Eq. \eqref{040}, we again expand the square root in Taylor series, but we hold terms to second order instead, to obtain
\begin{equation}
E(x,z)=\exp\left( -i \frac{\gamma_1}{k^3} z \right)  
\exp\left(  i\frac{\gamma_2}{k^3} z \, \hat{n} \right) 
\exp\left[i\frac{1}{2k^3} z \, \hat{n}^2  \right] E(x,0),
\end{equation}
where, for simplicity, we have defined
\begin{equation}
\gamma_1 \equiv  k^4-\frac{k^2}{2}-\frac{1}{8} , \qquad
\gamma_2 \equiv  k^2+\frac{1}{2}.
\end{equation}
We develop the initial condition in terms of the Gauss-Hermite functions
\begin{equation}
\varphi_m(x)=\left( \frac{1}{\pi}\right)^{1/4} \frac{1}{\sqrt{2^m m!}}\exp\left( -\frac{1}{2}x^2\right)H_m\left(x \right),
\end{equation}
where $H_m\left(x \right)$ are the Hermite polynomials, to obtain
\begin{align}\label{0100}
E(x,z)=\exp\left( -i \frac{\gamma_1}{k^3} z  \right)  
\exp\left(  i\frac{\gamma_2}{k^3}  z \, \hat{n} \right) 
\sum_{j=0}^{\infty}c_j   \exp\left[i\frac{1}{2k^3} z  j^2  \right] \varphi_j(x).
\end{align}
Now, for $z=l\pi k^3$ with $l$ any non-negative integer, we have
\begin{align}
E(x,z=l\pi k^3)=
\exp\left( - i l \gamma_1 \pi  \right)  
\exp\left( i l \gamma_2 \pi  \, \hat{n} \right) 
\left[ \sum_{j=0}^{\infty}c_{2j}  \varphi_{2j}(x)
+i^l \sum_{j=0}^{\infty}c_{2j+1}  \varphi_{2j+1}(x)\right] .
\end{align}
Next, considering the identities
\begin{equation}
2\sum_{j=0}^{\infty}c_{2j}  \varphi_{2j}(x)=
\sum_{j=0}^{\infty}c_{j}  \varphi_{j}(x)
+\left(-1 \right)^{ \hat{j}}
\sum_{j=0}^{\infty}c_{j}  \varphi_{j}(x)
=E(x,0)+E(-x,0)
\end{equation}
and
\begin{equation}
2\sum_{j=0}^{\infty}c_{2j+1}  \varphi_{2j+1}(x)=
\sum_{j=0}^{\infty}c_{j}  \varphi_{j}(x)
-\left(-1 \right)^{ \hat{j}}
\sum_{j=0}^{\infty}c_{j}  \varphi_{j}(x)
=E(x,0)-E(-x,0);
\end{equation}
we obtain
\begin{align}
E(x,z= l \pi k^3)=
\exp\left( -i l \gamma_1 \pi  \right)  
\big[
\frac{1+i^l}{2}\exp\left( i l \gamma_2 \pi  \, \hat{n}  \right) E(x,0)
+\frac{1-i^l}{2}\exp\left( i l \gamma_2 \pi  \, \hat{n}  \right) E(-x,0)
\big].
\end{align}
But, as we already said, $ \mathfrak{F}_\alpha \left\lbrace f(x) \right\rbrace =
 \exp\left(i \alpha \hat{n} \right) \left\lbrace  f(x)\right\rbrace  $ \cite{Fan}, thus
\begin{equation}\label{main}
E(x,z= l \pi k^3)=\exp\left( - i l \gamma_1 \pi  \right)  
\left[   \mathfrak{F}_{l\gamma_2 \pi } \left\lbrace \frac{1+i^l}{2} E(x,0)
+   \frac{1-i^l}{2} E(-x,0)\right\rbrace\right] .
\end{equation}
Hence, at these periodic distances the field is the fractional Fourier transform of a superposition of the initial field and its specular image. It is clear that if the initial condition is symmetric, $E(-x,0)=E(x,0)$, then at those periodic distances, we will have just the fractional Fourier transform of it. In particular, when $l$ is congruent with 0 modulo 4, the field is the fractional Fourier transform of the initial condition times a phase factor. If $l$ is congruent with 1 or with 3 modulo 4, we get the fractional Fourier transform of a superposition of the initial field and its specular image. In the case of $l$ congruent with 2 modulo 4, we obtain a phase factor times the fractional Fourier transform of the specular image of the initial condition; of course, in this last case, if the initial condition is symmetric we will have just the fractional Fourier transform of the initial condition. \\

An interesting case is obtained when the dimensionless wave vector $k$ is chosen such that $l \pi \gamma_2=2m\pi$, where now $m$ is another positive integer. As  $ \mathfrak{F}_{2\pi m} $ is the identity operator, we get
\begin{equation}\label{0240}
E(x,z= l \pi k_c^3)=\exp\left( - i l \gamma_c \pi  \right)  
\left[ \frac{1+i^l}{2}  E(x,0)
+\frac{1-i^l}{2}E(-x,0)   \right],
\end{equation}
where $k_c=\left( 2m/l-1/2\right)^{1/2} $ and $\gamma_c=3/8+m\left(4m-3l \right)/l^2$.
Thus, for those periodic distances and values of $k$, we can obtain the revival of the initial condition (when $l$ is congruent with $0$ mod 4), a superposition of the initial condition and its specular image (when $l$ is congruent with $1$ or with $3$ mod 4) and the specular image of the initial condition (when $l$ is congruent with $2$ mod 4). 
A similar result occurs when $l \pi \gamma_2=\left( 2m+1\right) \pi$, where again $m$ is a positive integer, for which the fractional Fourier transform becomes the parity operator; but in this case $E(x,0)$, in Eq. \eqref{0240} is replaced by $E(-x,0)$, and vice versa.
The easiest situation is when we pick $l=1$, and then, $z=\pi\left(2m-\frac{1}{2} \right)^{3/2} $, $k=\sqrt{2m-\frac{1}{2}}$ and $\gamma_1=m\left(4m-3\right)+\frac{3}{8} $. Below, we will study the Bessel functions and the Airy function as initial conditions, and this particular case will be exemplified.

\section{A Bessel function as initial condition}
In the particular case of a Bessel function as initial condition, $E(x,0)=J_\nu(bx+a)$, where $\nu$ is non-negative integer, we know from the Appendix A its fractional Fourier transform, Eq \eqref{tffBessel}, thus
\begin{align}\label{bessel}
E&\left( x,z=l\pi k^3\right) =
\exp\left\lbrace i \left[ -\left(\gamma_1+\frac{1}{2}\gamma_2 \right)l \pi +\frac{1}{2}\left(x^2+\frac{b^2}{2} \right) \tan\left(l\pi\gamma_2 \right) \right]  \right\rbrace 
\sqrt{\sec\left(l \pi \gamma_2 \right) }\; \; \times
\nonumber \\ &
\left\lbrace \frac{1+i^l}{2}
J_\nu^{\left( 2\right)} \left[ x\,b\,\sec\left( l \pi \gamma_2\right)+a, \frac{b^2}{4}\tan\left(l \pi \gamma_2 \right) ;-i\right]
+\frac{1-i^l}{2}
J_\nu^{\left( 2\right)} \left[- x\,b\,\sec\left( l \pi \gamma_2\right)+a, \frac{b^2}{4}\tan\left(l \pi \gamma_2 \right) ;i\right]
\right\rbrace  .
\end{align}
where $J_\nu^{(2)}(\xi,\zeta;i)$ is the second order generalized Bessel function, defined in Eq. \eqref{0290} of Appendix A.\\
\begin{figure}  [htbp!]
	\centering
	\begin{subfigure}{.45\textwidth}
		\centering
		\includegraphics[scale=0.65]{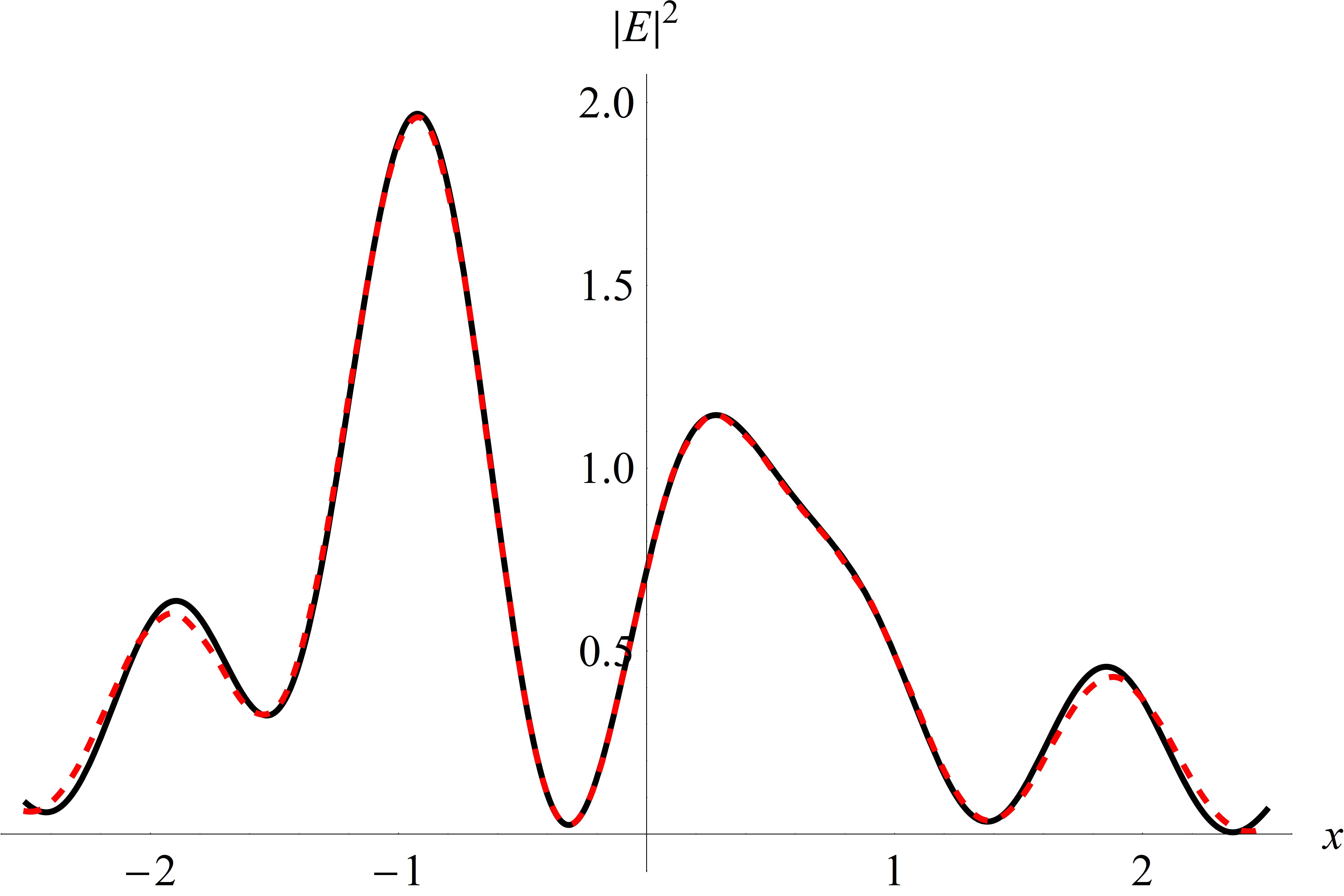}
		\caption{$\nu=0$}
	\end{subfigure}$\qquad$
	\begin{subfigure}{.45\textwidth}
		\centering
		\includegraphics[scale=0.65]{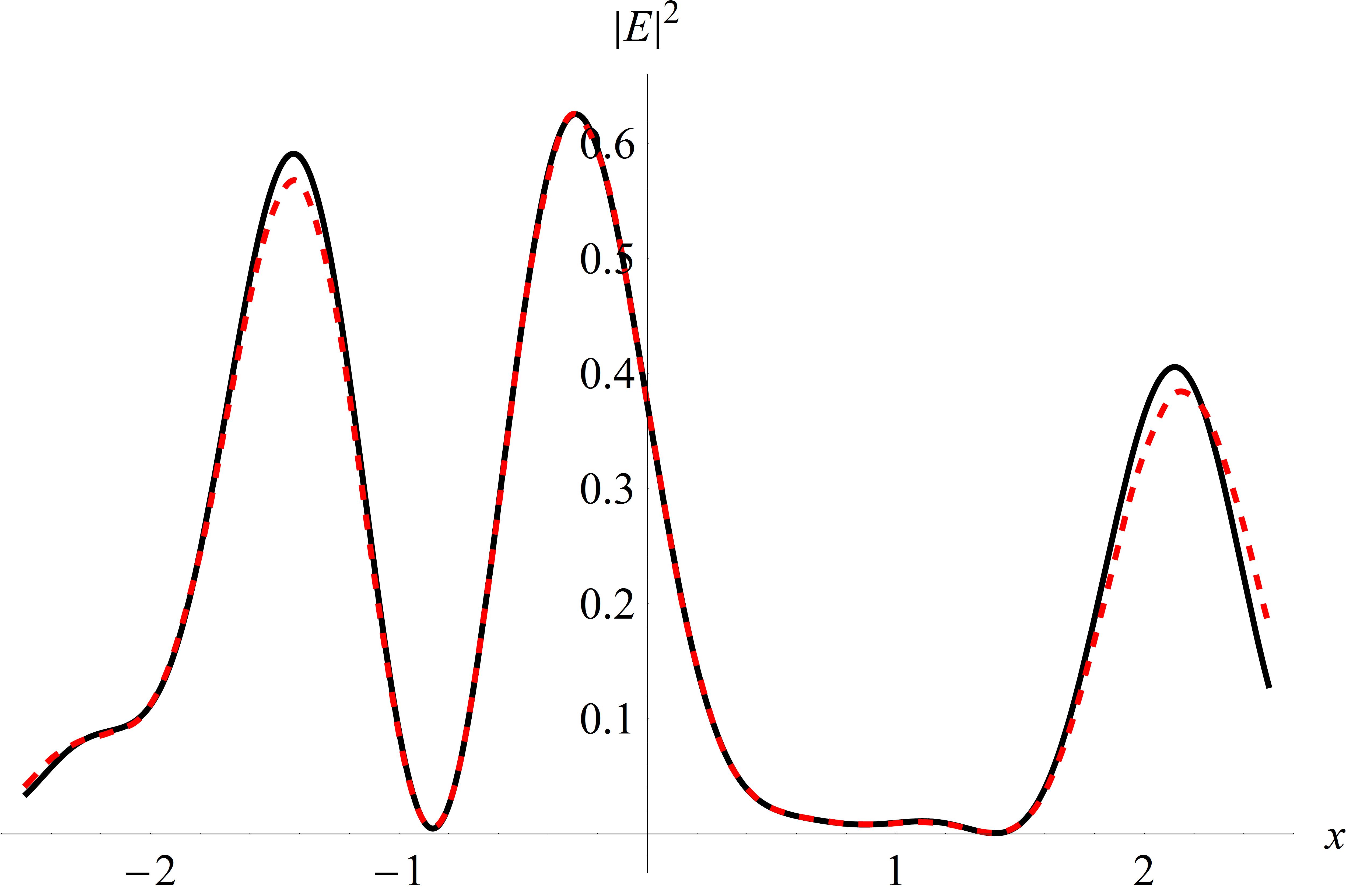}
		\caption{$\nu=3$}
	\end{subfigure}	
	\caption{The square of the propagated electric field at the first critical distance when the initial condition is $J_\nu\left( x+3\right) $. The quadratic GRIN media parameters are $n_0 = 1.5$ and $g = 10 \textrm{mm}^{-1}$. The beam has $k=1099.7$. The black continuous line is the graphic of expression \eqref{bessel} and the red dotted line is the exact numerical solution.}
		\label{figb1}
\end{figure}
In Figure \ref{figb1}, we show the field intensity at the first splitting distance $z =\pi k^3$, when the initial condition is a Bessel function $J_\nu (x+3)$. The parameters of the quadratic GRIN medium are $n_0 = 1.5$ and $g = 10 \textrm{mm}^{-1}$, and we have taken $k=1099.7$. The black continuous line is the  fractional Fourier transform given in Eq. \eqref{bessel} and the red dotted line is the exact numerical solution.
\begin{figure}  [htbp!]
	\centering
	\begin{subfigure}{.45\textwidth}
		\centering
		\includegraphics[scale=0.65]{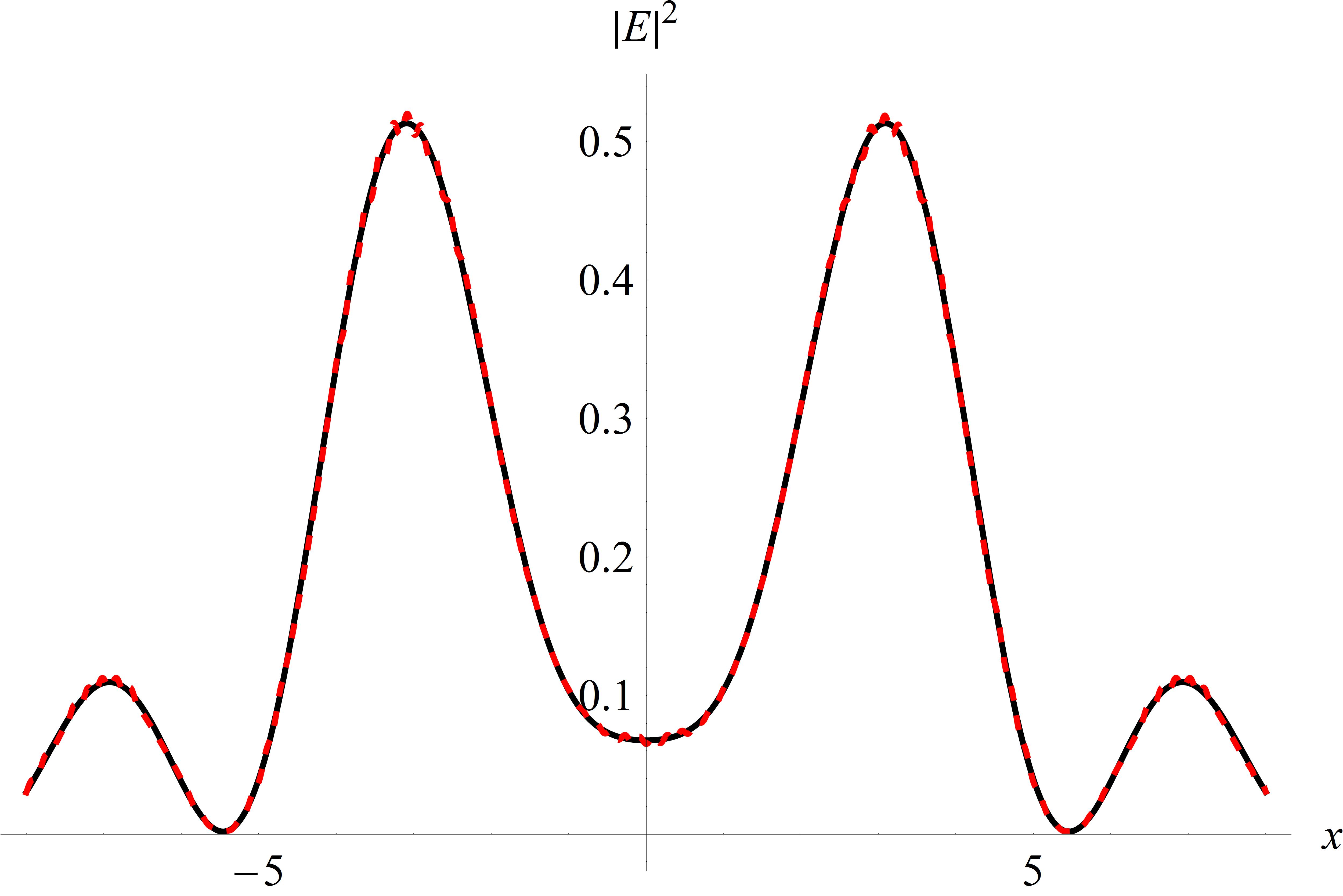}
		\caption{$\nu=0$}
	\end{subfigure}$\qquad$
	\begin{subfigure}{.45\textwidth}
		\centering
		\includegraphics[scale=0.65]{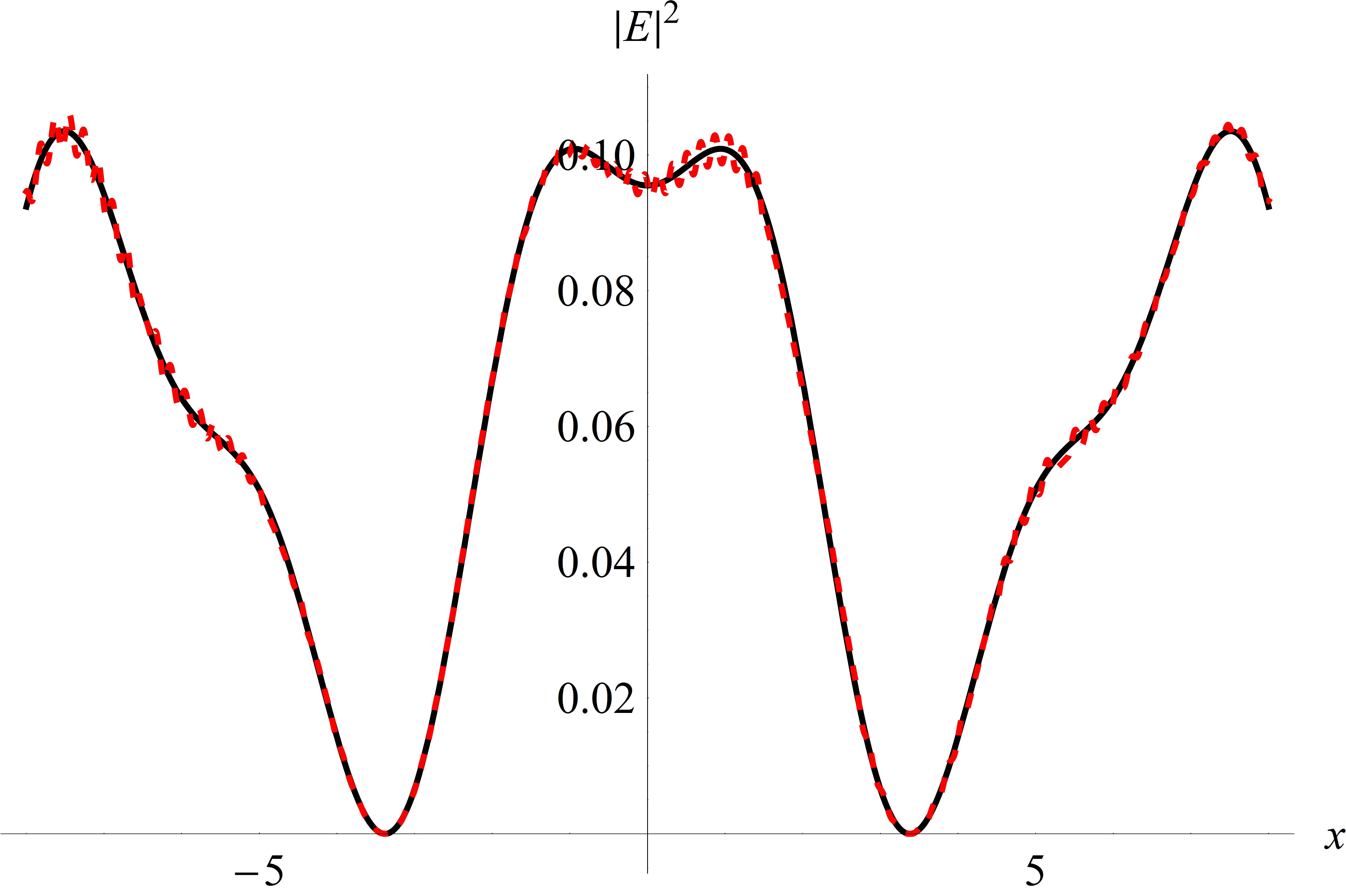}
		\caption{$\nu=3$}
	\end{subfigure}	
	\caption{The square of the propagated electric field at the first critical distance when the initial condition is $J_\nu\left( x+3\right)$. The quadratic GRIN media has $n_0 = 1.5$ and $g = 10 \textrm{mm}^{-1}$. We have taken $l=1$, $m=8\times10^5$ and then $k_c=1264.91$. The black continuous line is the graphic of expression \eqref{0240b} and the red dotted line is the exact numerical solution.}
		\label{figb2}
\end{figure}\\
For the special case indicated in Eq. \eqref{0240}, we have
\begin{equation}\label{0240b}
E(x,z= l \pi k_c^3)=\exp\left( - i l \gamma_c \pi  \right)  
\left[ \frac{1+i^l}{2} J_\nu (bx+a)
+\frac{1-i^l}{2}  J_\nu (-bx+a)   \right].
\end{equation}
In Fig. \ref{figb2}, we plot \eqref{0240b} (black continuous line) and the exact numerical solution (red dotted line), when $b=1$ and $a=3$. The GRIN media parameters are the same as in Fig. \ref{figb1}, and we took $l=1$ and $m=8\times10^5$, so $k_c=1264.91$. The result of the propagation is just a linear combination of the initial condition with its specular image.\\

\section{An Airy function as initial condition}
We take now as initial condition the Airy function $\mathrm{Ai}(bx+a)$. Considering the fractional Fourier transform of this initial condition, Eq. \eqref{fftairy} in Appendix B, the field propagated to the periodic distances $z=l\pi k^3$ is given by
\begin{align}\label{0190}
E & \left( x,z=l\pi k^3\right) =
\sqrt{\sec\left(l \pi \gamma_2 \right) }\;
\exp\left\lbrace i \left[ -\frac{1}{2} l \pi \gamma_2  +\frac{1}{2}\left( x^2-b^2a \right) \tan\left(l \pi \gamma_2 \right)+\frac{b^6}{12}\tan^3\left(l \pi \gamma_2 \right) \right] \right\rbrace 
\nonumber \\ &
\left\lbrace  \frac{1+i^l}{2}
\exp\left[ -i\frac{b^3}{2}x \tan\left(l \pi \gamma_2 \right)\sec\left(l \pi \gamma_2 \right)\right]
\mathrm{Ai}\left[ bx\sec\left(l \pi \gamma_2 \right)+a-\frac{b^4}{4}\tan^2\left(l \pi \gamma_2 \right)\right]
\right. \nonumber \\ &\left. 
+  \frac{1-i^l}{2}
\exp\left[ i\frac{b^3}{2}x \tan\left(l \pi \gamma_2 \right)\sec\left(l \pi \gamma_2 \right)\right]
\mathrm{Ai}\left[ -bx\sec\left(l \pi \gamma_2 \right)+a-\frac{b^4}{4}\tan^2\left(l \pi \gamma_2 \right)\right] \right\rbrace . 
\end{align}
\begin{figure}  [htbp!]
	\centering
	\includegraphics[scale=0.65]{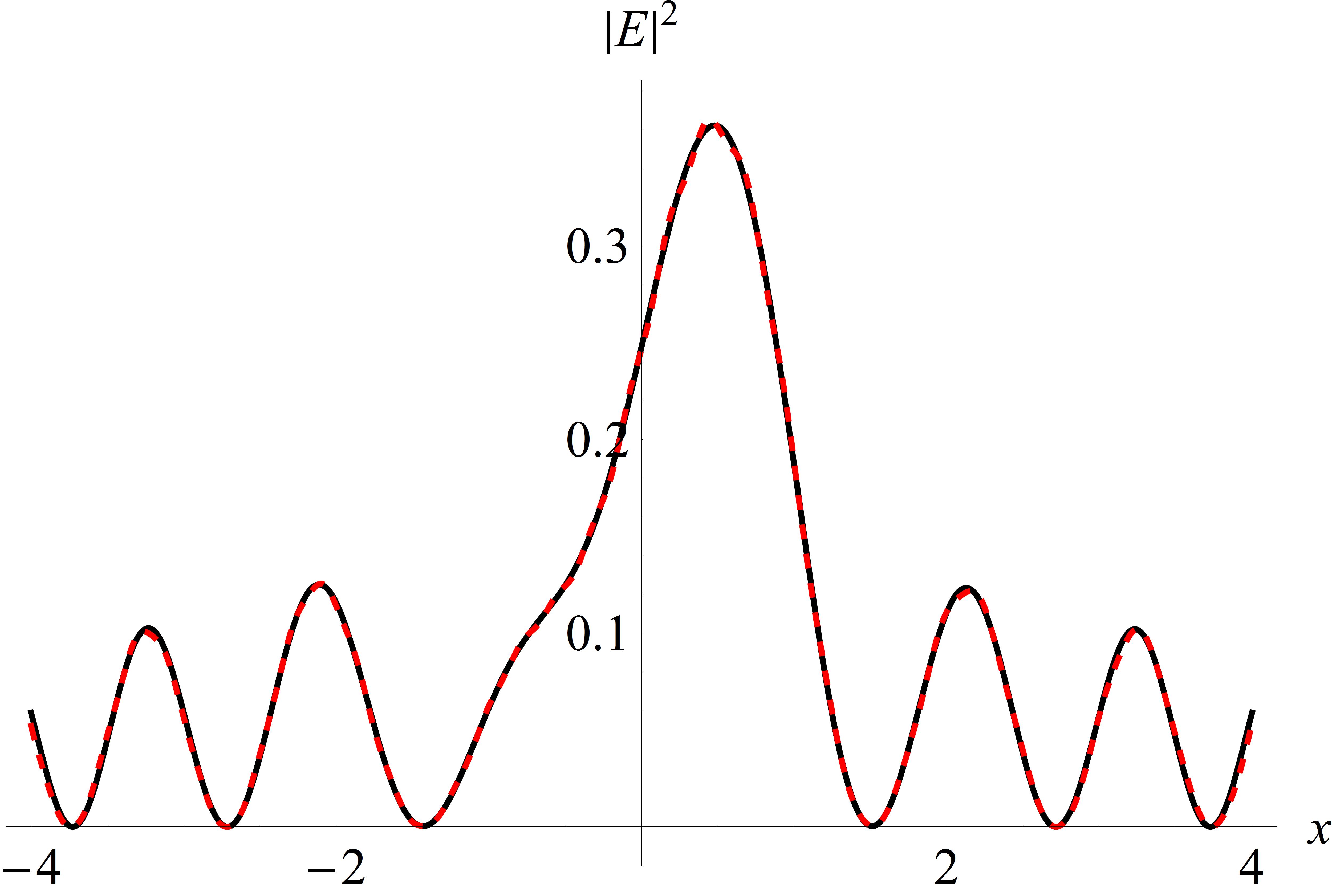}
	\caption{The square of the electric field at the critical distance when the initial condition is $\mathrm{Ai}\left(x \right) $. The quadratic GRIN media parameters are $n_0 = 1.5$ and $g = 10 \textrm{mm}^{-1}$, and $k=1000.5$. The black continuous line is the graphic of expression \eqref{0190} and the red dotted line is the exact numerical solution.}
	\label{figa1}
\end{figure}
In Figure \ref{figa1}, we show the field intensity at the first splitting distance $z =\pi k^3$, when the initial condition is an Airy function $\textrm{Ai}(x)$. The parameters of the quadratic GRIN medium are $n_0 = 1.5$ and $g = 10 \textrm{mm}^{-1}$, and we have taken $k=1000.5$. The black continuous line is the  fractional Fourier transform given in Eq. \eqref{0190} and the red dotted line is the exact numerical solution.\\
\begin{figure}  [htbp!]
	\centering
	\includegraphics[scale=0.65]{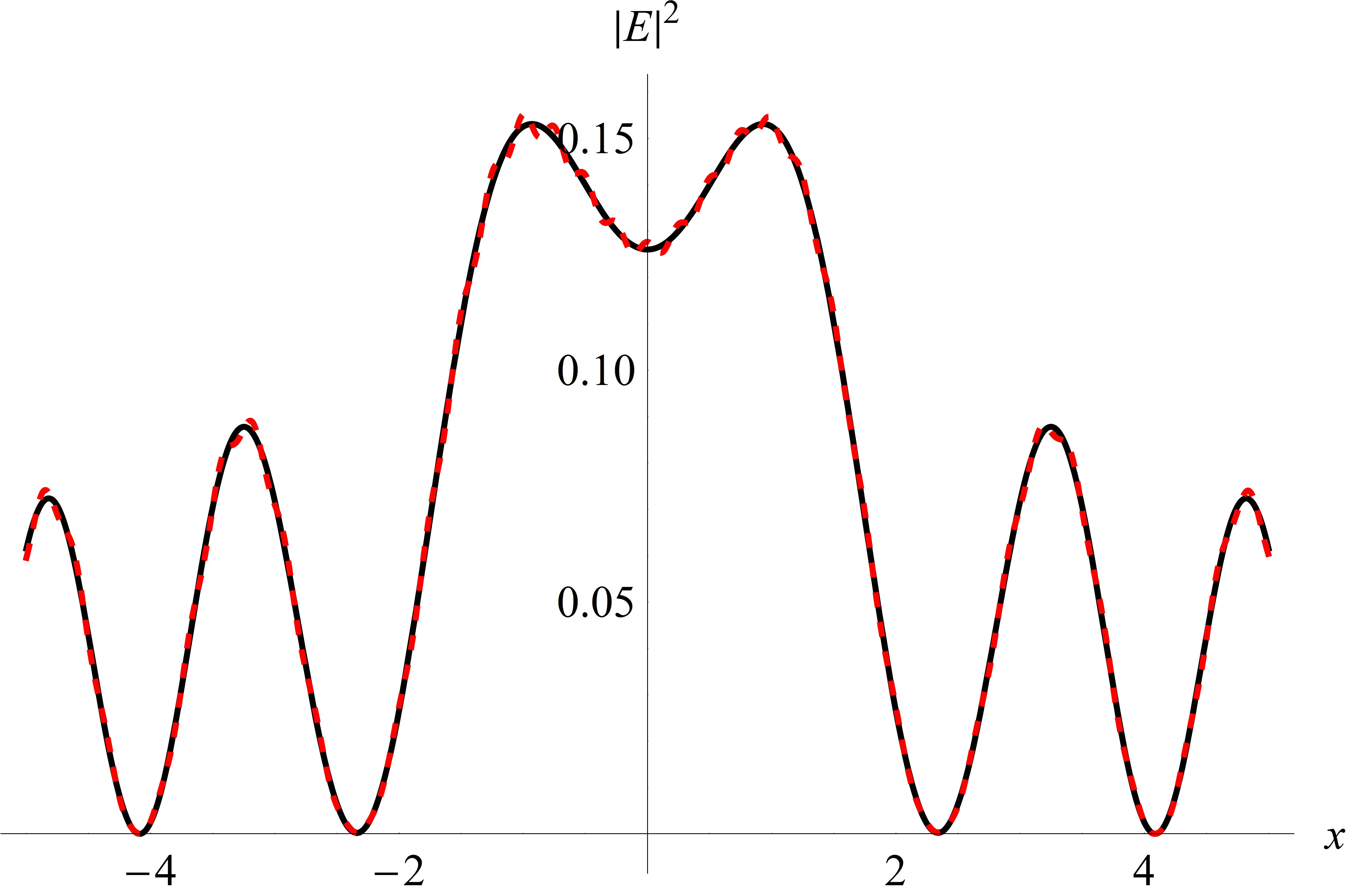}
	\caption{The square of the propagated electric field at the critical distance when the initial condition is $\mathrm{Ai}\left( x\right) $. The quadratic GRIN media has $n_0 = 1.5$ and $g = 10 \textrm{mm}^{-1}$. We have taken $l=1$, $m=4 \times 10^5$ and then $k=894.43$. The black continuous line is the graphic of expression \eqref{0240Ai} and the red dotted line is the exact numerical solution.}
	\label{figa2}
\end{figure}
For the special case indicated in Eq. \eqref{0240}, we have
\begin{equation}\label{0240Ai}
E(x,z= l \pi k_c^3)=\exp\left( - i l \gamma_c \pi  \right)  
\left[ \frac{1+i^l}{2} \textrm{Ai}(x+1)
+\frac{1-i^l}{2}  \textrm{Ai}(-x+1)   \right].
\end{equation}
In Fig. \ref{figa2}, we plot \eqref{0240Ai} (black continuous line) and the exact numerical solution (red dotted line). The GRIN media parameters are the same than in Fig. \ref{figa1}, and we took $l=1$ and $m=4 \times 10^5$, so $k=894.43$. The result of the propagation is just a linear combination of the initial condition with its specular image.

\section{Conclusions}
We have shown that the propagation of an initial field distribution in a quadratic GRIN media, {\it beyond the paraxial wave approximation}, produces the revival and the splitting of the input field, at specific propagation distances. It is also proved that the field at another propagation distances  is given by the fractional Fourier transform of the superposition of the initial field with a reflected version of it. We have applied the results to an initial Bessel function and found that the propagated field is given by a superposition of Generalized Bessel functions \cite{Dattoli,Torre}. Also, the example when the initial field is an Airy function is examined. In both concrete cases, our predictions are checked against the exact numerical solution and good agreement has been established.

\appendix

\section{The fractional Fourier transform of a Bessel function of the first kind of integer order}
It is known \cite{pellat} that the fractional Fourier transform $\mathfrak{F}_{\alpha } \{f\}$ of an arbitrary function $f$ is given by, $ \mathfrak{F}_{\alpha } \{f\}=e^{i \alpha  \hat{n}} \{f\} $, where $ \hat{n} $ is the number operator of the quantum harmonic oscillator.\\
To find the fractional Fourier transform of an integer order Bessel function of the first kind, we use the Jacobi-Angers integral representation \cite{abramowitz,nist}
\begin{equation}
J_n(bx+a)=\frac{1}{2 \pi }\underset{-\pi }{\overset{\pi }{\int }}d\tau \exp \{i\left[ n \tau -\left( bx+a \right)   \sin(\tau) \right] \}.
\end{equation}
Then we have,
\begin{align}
\mathfrak{F}_{\alpha } \left\{J_n(bx+a)\right\}&=
e^{i \alpha  \hat{n}}\frac{1}{2 \pi }
\int_{-\pi}^{\pi}d\tau
\exp \{i\left[ n \tau -\left( bx+a \right)   \sin(\tau) \right] \}
\nonumber\\
&=
\frac{1}{2 \pi }\int_{-\pi}^{\pi}d\tau
\exp\left(i n \tau \right) 
\exp\left( i \alpha  \hat{n}\right) 
\exp\left[-i \left( bx+a\right)  \sin(\tau) \right] .
\end{align}
We now write the number operator as $ \hat{n}=\frac{\hat{p}^2}{2}+\frac{x^2}{2}-\frac{1}{2} $, and so $\exp\left(i \alpha  \hat{n}\right)=\exp\left(-i \frac{\alpha}{2} \right) \exp\left[i \frac{\alpha}{2}\left( \hat{p}^2+x^2\right)  \right] $. But we know that \cite{Agarwal}
\begin{equation}
\exp\left[i \zeta \left( \hat{p}^2+x^2\right)  \right]=
\exp\left[i f(\zeta)x^2\right] 
\exp\left[-i g(\zeta) \left(x \hat{p}+\hat{p} x \right) \right] 
\exp\left[i f(\zeta)\hat{p}^2\right],
\end{equation}
where $f(\zeta)=\frac{1}{2}\tan(2\zeta)$ and $g(\zeta)=\frac{1}{2}\ln\left[\cos(2\zeta)\right] $. Thus,
\begin{align}
\mathfrak{F}_{\alpha } \left\{J_n(bx+a)\right\}=&
\frac{\exp\left(-i \frac{\alpha}{2} \right)\exp\left[i f\left( \frac{\alpha}{2}\right) x^2\right] }{2 \pi }\int_{-\pi}^{\pi}d\tau
\exp\left(i n \tau \right) 
\exp\left[-i g\left( \frac{\alpha}{2}\right)  \left(x \hat{p}+\hat{p} x \right) \right] 
\nonumber\\& 
\exp\left[i f\left( \frac{\alpha}{2}\right) \hat{p}^2\right]
\exp\left[-i \left( bx+a\right)  \sin(\tau) \right].
\end{align}
But,
\begin{equation}
\exp\left[i f\left( \frac{\alpha}{2}\right) \hat{p}^2\right]
\exp\left[-i \left( bx+a\right)  \sin(\tau) \right]=
\exp\left[i f\left( \frac{\alpha}{2}\right)b^2 \sin^2\left( \tau\right) \right]
\exp\left[-i \left( bx+a\right)  \sin(\tau) \right],
\end{equation}
then, we have
\begin{align}
\mathfrak{F}_{\alpha } \left\{J_n(bx+a)\right\}=&
\frac{\exp\left(-i \frac{\alpha}{2} \right) \exp\left[i f\left( \frac{\alpha}{2}\right) x^2\right]}{2 \pi }\int_{-\pi}^{\pi}d\tau
\exp\left(i n \tau \right) 
\exp\left[i f\left( \frac{\alpha}{2}\right)b^2 \sin^2\left(\tau \right) \right]
\nonumber\\& 
\exp\left[-i g\left( \frac{\alpha}{2}\right)  \left(x \hat{p}+\hat{p} x \right) \right] 
\exp\left[-i \left( bx+a\right)  \sin(\tau) \right] .
\end{align}
Also
\begin{equation}
\exp\left[-i g\left( \frac{\alpha}{2}\right)  \left(x \hat{p}+\hat{p} x \right) \right] 
\exp\left[-i \left( bx+a\right)  \sin(\tau) \right]
=\exp\left[ -g\left( \frac{\alpha}{2}\right) \right]
\exp\left\lbrace  -i  \left( \exp\left[-2 g\left(\frac{\alpha}{2} \right)  \right] bx+a\right)  \sin\left( \tau\right)\right\rbrace, 
\end{equation}
then the fractional Fourier transform of the Bessel functions of the first kind can be written as
\begin{align}
\mathfrak{F}_{\alpha }& \left\{J_n(bx+a)\right\}=
\exp\left(-i \frac{\alpha}{2} \right) \exp\left[i f\left( \frac{\alpha}{2}\right) x^2\right]
\exp\left[ -g\left( \frac{\alpha}{2}\right) \right]
\nonumber\\& 
\times \frac{1}{2 \pi }
\int_{-\pi}^{\pi}d\tau
\exp\left(i n \tau \right) 
\exp\left[i f\left( \frac{\alpha}{2}\right)b^2 \sin^2\left(\tau \right) \right]
\exp\left\lbrace  -i  \left(\exp\left[-2 g\left(\frac{\alpha}{2} \right)  \right] bx+a\right)  \sin\left( \tau\right)\right\rbrace .
\end{align}
Writing
$   \sin ^2\left( \tau\right) =\frac{1}{2}\left[ 1- \cos (2 \tau )\right] $
and changing the integration variable from $\tau$ to $-\tau$ in the last formula, we arrive to
\begin{align}\label{0170}
&\mathfrak{F}_{\alpha } \left\{J_n(bx+a)\right\}=
\exp\left[ -g\left( \frac{\alpha}{2}\right) \right] \exp\left[-i \frac{\alpha}{2}+i f\left( \frac{\alpha}{2}\right) x^2+\frac{i}{2}b^2 f\left( \frac{\alpha}{2}\right) 
	\right]
\nonumber\\&
\frac{1}{2 \pi}
\int_{-\pi}^{\pi}d\tau
\exp\left(-i n \tau \right) 
\exp\left[  -  \frac{i}{2} f\left( \frac{\alpha}{2}\right) b^2 \cos\left(2\tau \right) \right] 
\exp\left\lbrace  i  \left(\exp\left[-2 g\left(\frac{\alpha}{2} \right)  \right] bx+a\right)  \sin\left( \tau\right)\right\rbrace .
\end{align}
Introducing now the Generalized Bessel Functions, defined as \cite{Dattoli}
\begin{equation}\label{0290}
J_n^{(m)}\left(x,y;c \right) =\sum_{l=-\infty}^{\infty}c^l J_{n-ml} \left( x\right) J_l\left(y \right) ,
\end{equation}
and its integral representation
\begin{equation}
J_n^{(m)}\left(x,y;e^{i\theta} \right) =\frac{1}{2\pi}\int_{-\infty}^{\infty}
d\varphi \exp\left[  ix\sin\varphi+iy\sin\left( m\varphi+\theta\right)-in\varphi  \right] ,
\end{equation}
we can cast   \eqref{0170} as
\begin{align}
\mathfrak{F}_{\alpha } \left\{J_n(bx+a)\right\}=
\exp\left[-g\left( \frac{\alpha}{2}\right)-i \frac{\alpha}{2}+i f\left( \frac{\alpha}{2}\right) \left( x^2+\frac{b^2}{2}\right) \right] \, 
J_n^{(2)} \left[ e^{-2g\left( \frac{\alpha}{2}\right) }\,bx+a,\frac{b^2}{2}f\left(\frac{\alpha}{2} \right) ;-i \right] .
\end{align}
Substituting the functions $f(\zeta)=\frac{1}{2}\tan(2\zeta)$ and $g(\zeta)=\frac{1}{2}\ln\left[\cos(2\zeta)\right] $, we finally obtain
\begin{align}\label{tffBessel}
\mathfrak{F}_{\alpha } \left\{J_n(bx+a)\right\}
=\sqrt{\sec \alpha } \;
\exp\left\lbrace  \frac{i}{2}\left[-\alpha+  \left( x^2+\frac{b^2}{2} \right) \tan \alpha \right] 
\right\rbrace   \, 
J_n^{(2)}   \left(   b \, x \sec\alpha + a ,\frac{b^2}{4}\tan \alpha ; -i  \right)   .
\end{align}

\section{The fractional Fourier transform of an Airy function}
We use again the known fact \cite{pellat} that the fractional Fourier transform  $\mathfrak{F}_{\alpha } \{f\}$ of an arbitrary function $f$ is given by, $ \mathfrak{F}_{\alpha } \{f\}=e^{i \alpha  \hat{n}} \{f\} $ where $ \hat{n} $ is the number operator of the quantum harmonic oscillator.\\
To find the fractional Fourier transform of an Airy function, we use the integral representation \cite{abramowitz,nist}
\begin{equation}\label{ftairy}
\mathrm{Ai}(bx+a)=\frac{1}{2 \pi }
\underset{-\infty }{\overset{\infty }{\int }} d\tau \exp{\left[  i \tau \left( bx+a\right)  \right]  }   \exp{\left(i \frac{\tau^3}{3} \right)} .
\end{equation}
Then we have,
\begin{align}
\mathfrak{F}_{\alpha } \left\{\mathrm{Ai}(bx+a)\right\}&=
e^{i \alpha  \hat{n}}
\frac{1}{2 \pi }
\underset{-\infty }{\overset{\infty }{\int }} d\tau \exp{\left( i \tau b x \right) }  \exp{\left( i\tau a\right) } \exp{\left(i \frac{\tau^3}{3} \right)}
\nonumber\\ &
=
\frac{1}{2 \pi }
\underset{-\infty }{\overset{\infty }{\int }} d\tau    \exp{\left(i \frac{\tau^3}{3} \right)}
\exp{\left( i\tau a\right) } \;   e^{i \alpha  \hat{n}}
\exp{\left( i \tau b x \right) } .
\end{align}
We now write the number operator as $ \hat{n}=\frac{\hat{p}^2}{2}+\frac{x^2}{2}-\frac{1}{2} $, and so $\exp\left(i \alpha  \hat{n}\right)=\exp\left(-i \frac{\alpha}{2} \right) \exp\left[i \frac{\alpha}{2}\left( \hat{p}^2+x^2\right)  \right] $. But we know that \cite{Agarwal}
\begin{equation}
\exp\left[i \zeta \left( \hat{p}^2+x^2\right)  \right]=
\exp\left[i f(\zeta)x^2\right] 
\exp\left[-i g(\zeta) \left(x \hat{p}+\hat{p} x \right) \right] 
\exp\left[i f(\zeta)\hat{p}^2\right],
\end{equation}
where $f(\zeta)=\frac{1}{2}\tan(2\zeta)$ and $g(\zeta)=\frac{1}{2}\ln\left[\cos(2\zeta)\right] $. Thus,
\begin{align}
\mathfrak{F}_{\alpha } \left\{\mathrm{Ai}(bx+a)\right\}=&
\frac{\exp\left(-i \frac{\alpha}{2} \right)\exp\left[i f\left( \frac{\alpha}{2}\right) x^2\right] }{2 \pi }\int_{-\infty}^{\infty}d\tau
\exp\left(i \frac{\tau^3}{3} \right) 
\exp\left[-i g\left( \frac{\alpha}{2}\right)  \left(x \hat{p}+\hat{p} x \right) \right] 
\exp{\left( i\tau a\right) }
\nonumber\\& 
\times  \exp\left[i f\left( \frac{\alpha}{2}\right) \hat{p}^2\right]
\exp\left( i \tau b x \right) .
\end{align}
But,
\begin{equation}
\exp\left[i f\left( \frac{\alpha}{2}\right) \hat{p}^2\right]
\exp\left( i \tau b x \right)=
\exp\left[i f\left( \frac{\alpha}{2}\right)b^2 \tau^2 \right]
\exp\left( i \tau b x \right),
\end{equation}
then, we have
\begin{align}
\mathfrak{F}_{\alpha } \left\{\mathrm{Ai}(bx+a)\right\}=&
\frac{\exp\left(-i \frac{\alpha}{2} \right) \exp\left[i f\left( \frac{\alpha}{2}\right) x^2\right]}{2 \pi }\int_{-\infty}^{\infty}d\tau
\exp\left(i  \frac{\tau^3}{3} \right) 
\exp\left[i f\left( \frac{\alpha}{2}\right)b^2 \tau^2 \right]
\exp{\left( i\tau a\right) }
\nonumber\\& 
\times \exp\left[-i g\left( \frac{\alpha}{2}\right)  \left(x \hat{p}+\hat{p} x \right) \right] 
\exp\left( i b x \tau \right)  .
\end{align}
Also
\begin{equation}
\exp\left[-i g\left( \frac{\alpha}{2}\right)  \left(x \hat{p}+\hat{p} x \right) \right] 
\exp\left( i b x \tau\right) 
=\exp\left[ -g\left( \frac{\alpha}{2}\right) \right]
\exp\left\lbrace  i \exp\left[-2 g\left(\frac{\alpha}{2} \right)  \right] b x  \tau\right\rbrace, 
\end{equation}
then the fractional Fourier transform of the Airy function can be written as
\begin{align}
\mathfrak{F}_{\alpha }& \left\{\mathrm{Ai}(bx+a)\right\}=
\exp\left(-i \frac{\alpha}{2} \right) \exp\left[i f\left( \frac{\alpha}{2}\right) x^2\right]
\exp\left[ -g\left( \frac{\alpha}{2}\right) \right]
\nonumber\\& 
\times \frac{1}{2 \pi }
\int_{-\infty}^{\infty}d\tau
\exp\left(i  \frac{\tau^3}{3} \right)
\exp{\left( i\tau a\right) }
\exp\left[i f\left( \frac{\alpha}{2}\right)b^2 \tau^2  \right]
\exp\left\lbrace  i \, \exp\left[-2 g\left(\frac{\alpha}{2} \right)  \right] b x \tau\right\rbrace .
\end{align}
Completing a cube binomial and changing the integration variable, we get
\begin{align}
\mathfrak{F}_{\alpha } \left\{\mathrm{Ai}\left( b x+a\right) \right\}&=
\exp\left\lbrace -i \frac{\alpha}{2} 
+i f\left( \frac{\alpha}{2}\right) x^2
+i \frac{2}{3}b^6 f^3\left(\frac{\alpha}{2} \right)
-i b^3 f\left( \frac{\alpha}{2}\right) \exp\left[ -2 g\left( \frac{\alpha}{2}\right) \right] x
-g\left( \frac{\alpha}{2}\right)
-i b^2 f\left( \frac{\alpha}{2}\right) a 
\right\rbrace  
\nonumber\\& 
\times \; \frac{1}{2 \pi }
\int_{-\infty}^{\infty}d\tau
\exp\left(i  \frac{\tau^3}{3} \right)
\exp{\left( i\tau a\right) }
\exp\left(  i \tau \left\lbrace 
\exp\left[-2 g\left(\frac{\alpha}{2} \right)  \right] b x
-b^4f^2 \left(\frac{\alpha}{2} \right)
\right\rbrace 
\right) .
\end{align}
Finally, recalling the integral representation of the Airy function \eqref{ftairy} and the definitions of the functions $f$ and $g$, we obtain
\begin{align}\label{fftairy}
\mathfrak{F}_{\alpha } \left\{\mathrm{Ai}(bx+a)\right\}=&
\sqrt{\sec \alpha  } \;
\exp\left\lbrace i  \left[  -\frac{\alpha}{2}
+\frac{1}{2}\tan \alpha  \left(  x^2   -b^2 a
- x \, b^3 \sec\alpha 
+\frac{b^6}{6}\tan^2 \alpha 
\right) \right]  \right\rbrace
\nonumber\\&
\times \; 
\mathrm{Ai}\left( x \, b \sec \alpha +a
-\frac{b^4}{4} \tan^2  \alpha \right). 
\end{align}


\begin{thebibliography}{99}
\bibitem{gomez} C. Gomez-Reino, M.V. Perez, and C. Bao, \textit{Gradient-Index Optics} (Springer-Verlag, 2002).
\bibitem{R1} Enrique Silvestre-Mora, Pedro Andres and Jorge Ojeda-Castaneda,
\textquotedblleft Self-imaging in GRIN media,\textquotedblright \; Proc. SPIE 2730, Second Iberoamerican Meeting on Optics, 468 (February 5, 1996); doi:10.1117/12.231121.
\bibitem{R2} M. T. Flores-Arias, C. Bao, M. V. Pérez, and C. Gómez-Reino, \textquotedblleft Talbot effect in a tapered gradient-index medium for nonuniform and uniform illumination,\textquotedblright \; J. Opt. Soc. Am. A 16, 2439-2446 (1999).
\bibitem{ozaktas} H. M. Ozaktas and D. Mendlovic,\textquotedblleft Fractional Fourier transforms and their optical implementation. II,\textquotedblright \; J. Opt. Soc. Am. A 10, 2522-2531 (1993).
\bibitem{ojeda} Jorge Ojeda-Casta\~neda and Pete Szwaykowski, \textquotedblleft Novel modes in $\alpha$-power GRIN,\textquotedblright \; Proc. SPIE 1500, Innovative Optics and Phase Conjugate Optics, 246 (October 1, 1991); doi:10.1117/12.46835; http://dx.doi.org/10.1117/12.46835.
\bibitem{Arr} H. M. Moya-Cessa, M. Fern\'andez Guasti , V. M. Arrizon and S. Ch\'avez-Cerda,  \textquotedblleft Optical realization of quantum-mechanical invariants,\textquotedblright \; Opt. Lett. 34, 1459-1461 (2009).
\bibitem{Chuma} S. M. Chumakov and K. B. Wolf, \textquotedblleft Supersymmetry in Helmholtz optics,\textquotedblright \; Phys. Lett. A 193, 51-53 (1994).
\bibitem{David} A. Zu\~{n}iga-Segundo, B. M. Rodr\'iguez-Lara, D. J. Fern\'andez and H. M. Moya-Cessa , \textquotedblleft Jacobi photonic lattices and their SUSY partners,\textquotedblright \; Opt. Express 22, 987-994 (2014).
\bibitem{Agarwal} G. S. Agarwal and R. Simon, \textquotedblleft A simple realization of fractional Fourier transform and relation to harmonic oscillator Green's function,\textquotedblright \; Opt. Commun. 110, 23-26 (1994).
\bibitem{Fan} H. Y. Fan and J. H. Chen, \textquotedblleft On the core of the fractional Fourier transform and its role in composing complex fractional Fourier transformations and Fresnel transformations,\textquotedblright \; Front. Phys. 10, 1-6 (2015).
\bibitem{Eguibar} F. Soto-Eguibar, V. Arrizon,  A. Zu\~{n}iga-Segundo and H. M. Moya-Cessa, \textquotedblleft Optical realization of quantum Kerr medium dynamics,\textquotedblright \; Opt. Lett. 39, 6158-6161 (2014).
\bibitem{Arrizon} V. Arrizon, F. Soto-Eguibar, A. Zu\~{n}iga-Segundo and H. M. Moya-Cessa, \textquotedblleft Revival and splitting of a Gaussian beam in
gradient index media,\textquotedblright \; J. of the Opt. Soc. of Am. A 32, 1140-1145 (2015).
\bibitem{Dattoli} G. Dattoli, L. Giannessi,  L. Mezi  and  A. Torre, \textquotedblleft Theory of generalized Bessel functions,\textquotedblright \; Il Nuovo Cimento B 105(3), 327-348 (1990).
\bibitem{Torre} G. Dattoli, A. Torre, S. Lorenzutta and G. Maino,\textquotedblleft A note on the theory of \textit{n}-variable generalized bessel functions,\textquotedblright \; Il Nuovo Cimento B 106(10), 1159-1166 (1991).
\bibitem{Dattoli2} G. Dattoli and A. Torre, \textquotedblleft Note on “discrete-like diffraction dynamics in free space”: highlighting the variety of solving procedures,\textquotedblright \; J. of the Opt. Soc. of Am. B 31, 2214-2220 (2014).
\bibitem{LeijaI} A. Perez-Leija, F. Soto-Eguibar, S. Chavez-Cerda, A. Szameit, H. M. Moya-Cessa H. and D. N. Christodoulides, \textquotedblleft Discrete-like diffraction dynamics in free space,\textquotedblright \; Optics Express 21, 17951-17960 (2013).
\bibitem{LeijaII} T. Eichelkraut, C. Vetter, A. Perez-Leija, H. M. Moya-Cessa, D. N. Christodoulides  and A. Szameit, \textquotedblleft Coherent random walks in free space,\textquotedblright \; Optica 1, 268-271 (2014).
\bibitem{Torre2} A. Torre, \textquotedblleft Propagating Airy wavelet-related patterns,\textquotedblright \; J. Opt. 17, 075604 (2015).
\bibitem{wolf} K. B. Wolf, \textit {Geometric optics in phase space} (Springer-Verlag, 2004).
\bibitem{namias} V. Namias, \textquotedblleft The Fractional Order Fourier Transform and its Application
to Quantum Mechanics,\textquotedblright \; J. Inst. Maths Applics. 25, 241-265 (1980).
\bibitem{pellat} Pierre Pellat-Finet,  \textit{Optique de Fourier. Th\'eorie m\'etaxiale et fractionnaire}  (Springer-Verlag, France 2009).
\bibitem{walls} H. P. Yuen ,\textquotedblleft Two-photon coherent states of the radiation field,\textquotedblright \; Phys . Rev . A 13, 2226-2243 (1976).
\bibitem{1}  E. Schr\"odinger, \textquotedblleft Die gegenwärtige Situation in der Quantenmechanik I,\textquotedblright \; Die Naturwissenschaften, 23, 807-812, \textquotedblleft Die gegenwärtige Situation in der Quantenmechanik II,\textquotedblright \; Die Naturwissenschaften, 23, 823-828, \textquotedblleft Die gegenwärtige Situation in der Quantenmechanik III,\textquotedblright \; Die Naturwissenschaften, 23, 844-849; (1935).
\bibitem{abramowitz} M. Abramowitz and I. A. Stegun \textit{Handbook of Mathematical Functions with Formulas, Graphs and Mathematical Tables} (1972).
\bibitem{nist}  F. W. Olver, D. W. Lozier, R. F. Boisvert and C. W. Clark,  \textit{NIST Handbook of Mathematical Functions}. NIST (2010) 
\end{thebibliography}
\end{document}